\documentclass[italian,english]{article}
\usepackage[latin9]{inputenc}
\usepackage{color}
\usepackage{graphicx}
\usepackage{amssymb}

\makeatletter
\newcommand{\lyxaddress}[1]{
\par {\raggedright #1
\vspace{1.4em}
\noindent\par}
}

\makeatother

\usepackage{babel}

\begin{document}

\title{\textbf{On the longitudinal response function of interferometers
for massive gravitational waves from a bimetric theory of gravity}}

\author{\textbf{Christian Corda}}

\maketitle

\lyxaddress{\begin{center}
Associazione Galileo Galilei, Via Pier Cironi 16 - 59100 PRATO, Italy
and Centro di Scienze Naturali, Via di Galceti 74 - 59100 PRATO, Italy
\par\end{center}}

\lyxaddress{\begin{center}
\textit{E-mail address:} \textcolor{blue}{christian.corda@ego-gw.it} 
\par\end{center}}

Recently, some papers in the literature have shown that, from a bimetric
theory of gravity, it is possible to produce massive gravitational
waves which generate a longitudinal component in a particular polarization
of the wave. After a review of previous works, in this paper the longitudinal
response function of interferometers for this particular polarization
of the wave is computed in two different gauges, showing the gauge
invariance, and in its full frequency dependence, with specific application
to the Virgo and LIGO interferometers.

\section{Introduction}

The data analysis of interferometric gravitational waves (GWs) detectors
has recently started (for the current status of GWs interferometers
see \cite{key-1,key-2,key-3,key-4,key-5,key-6,key-7,key-8}) and the
scientific community hopes in a first direct detection of GWs in next
years. 

Detectors for GWs will be important for a better knowledge of the
Universe and also to confirm or ruling out the physical consistency
of General Relativity or of any other theory of gravitation \cite{key-9,key-10,key-11,key-12,key-13,key-14,key-15}.
This is because, in the context of Extended Theories of Gravity, some
differences between General Relativity and the others theories can
be pointed out starting by the linearized theory of gravity \cite{key-9,key-10,key-12,key-14,key-15}.
In this picture, recently, some papers in the literature have shown
that, from a bimetric theory of gravity, it is possible to produce
massive gravitational waves which generate a longitudinal component
in a particular polarization of the wave \cite{key-14,key-15}. After
a review of previous works (i.e. the work of de Paula, Miranda and
Marinho \cite{key-15} and my previous research \cite{key-14}) on
this topic, which is due to provide a context to bring out the relevance
of the results, in this paper the longitudinal response function of
interferometers for this particular polarization of the wave is computed
in its full frequency dependence and in two different gauges, showing
the gauge invariance, with specific application to the Virgo and LIGO
interferometers.

\section{A review of previous results on massive gravitational waves from
the bimetric theory of gravity}

An extension of linearized general relativity which takes into account
massive gravitons gives a weak-field stress-energy tensor \cite{key-14,key-15}

\begin{equation}
T_{\mu\nu}^{(m)}=-\frac{m_{g}}{8\pi}\{h_{\mu\nu}-\frac{1}{2}[(g_{0}^{-1})^{\alpha\beta}h_{\alpha\beta}](g_{0})_{\mu\nu}\},\label{eq: TEI}\end{equation}

where $m_{g}$ is the mass of the graviton, and $(g_{0})_{\mu\nu}$
the non-dynamical background metric (note: differently from \cite{key-15}
in this paper we work with $G=1$, $c=1$ and $\hbar=1$, exactly
like in \cite{key-14}). In this way the field equations can be obtained
in an einstenian form like

\begin{equation}
G_{\mu\nu}=-8\pi(T_{\mu\nu}+T_{\mu\nu}^{(m)}),\label{eq: einstein}\end{equation}

where $T_{\mu\nu}$ is the ordinary stress-energy tensor of the matter.
General relativity is recovered in the limit $m_{g}\rightarrow0$.

Calling $g{}_{\mu\nu}$ the dynamic metric and putting 

\begin{equation}
g_{\mu\nu}=\eta_{\mu\nu}+h_{\mu\nu}\label{eq: linearizza}\end{equation}

with $|h_{\mu\nu}|\ll1$ equation (\ref{eq: einstein}) can be linearized
in vacuum (i.e. $T_{\mu\nu}=0$) obtaining

\begin{equation}
{}\square\overline{h}_{\mu\nu}=m_{g}^{2}\overline{h}_{\mu\nu},\label{eq: linearizzata}\end{equation}

where $\square$ is the d'Alembertian operator and $\overline{h}_{\mu\nu}\equiv h_{\mu\nu}-\frac{h}{2}\eta_{\mu\nu}.$

The general solution of this equation is \cite{key-14,key-15}

\begin{equation}
\overline{h}_{\mu\nu}=e_{\mu\nu}\exp(ik^{\alpha}x_{\alpha}),\label{eq: sol S}\end{equation}

where $e_{\mu\nu}$ is the polarization tensor. 

The condition of normalization $k^{\alpha}k_{\alpha}=m_{g}^{2}$ gives
$k=\sqrt{\omega^{2}-m_{g}^{2}}$ and a speed of propagation 

\begin{equation}
v(\omega)=\frac{\sqrt{\omega^{2}-m_{g}^{2}}}{\omega},\label{eq: velocita' di gruppo 2}\end{equation}

which is exactly the velocity of a massive particle with mass $m_{g}$
(it is also the group-velocity of a wave-packet \cite{key-12,key-14,key-16,key-17,key-18}).

Thus, assuming that the wave is propagating in the $z$ direction,
the metric perturbation (\ref{eq: sol S}) can be rewritten like 

\begin{equation}
\overline{h}_{\mu\nu}=e_{\mu\nu}\exp(ikz-i\omega).\label{eq: sol S 2}\end{equation}

Using a tetrade formalism, the authors of \cite{key-15} found six
independent polarizations states (see equations 28-33 of \cite{key-15}),
while in \cite{key-14} it has been shown that, from the polarization
labelled with $\Phi_{22}$ in \cite{key-15} (equations 32 and 38
of \cite{key-15}), a longitudinal force is present.

In fact, let us consider equation 38 of \cite{key-15}. Putting $h_{g}\equiv h_{00}+h_{33}$,
this equation can be rewritten as \cite{key-14}

\begin{equation}
\Phi_{22}=\frac{1}{8}h_{g}(t-vz).\label{eq: hg}\end{equation}

Taken in to account only the $\Phi_{22}$ polarization in equation
(\ref{eq: sol S}) one gets \cite{key-14}\begin{equation}
\overline{h}_{\mu\nu}(t,z)=\frac{1}{8}h_{g}(t-vz)\eta_{\mu\nu}\label{eq: perturbazione scalare}\end{equation}
and the corrispondent line element is the conformally flat one \cite{key-14}

\begin{equation}
ds^{2}=[1+\frac{1}{8}h_{g}(t-vz)](-dt^{2}+dz^{2}+dx^{2}+dy^{2}).\label{eq: metrica puramente scalare}\end{equation}
Because the analysis on the motion of test masses is performed in
a laboratory environment on Earth, the coordinate system in which
the space-time is locally flat is typically used and the distance
between any two points is given simply by the difference in their
coordinates in the sense of Newtonian physics \cite{key-12,key-14,key-16,key-17,key-19}.
This frame is the proper reference frame of a local observer, located
for example in the position of the beam splitter of an interferometer.
In this frame gravitational waves manifest themself by exerting tidal
forces on the masses (the mirror and the beam-splitter in the case
of an interferometer). A detailed analysis of the frame of the local
observer is given in ref. \cite{key-19}, sect. 13.6. Here only the
more important features of this coordinate system are recalled:

the time coordinate $x_{0}$ is the proper time of the observer O;

spatial axes are centered in O;

in the special case of zero acceleration and zero rotation the spatial
coordinates $x_{j}$ are the proper distances along the axes and the
frame of the local observer reduces to a local Lorentz frame: in this
case the line element reads \cite{key-19}

\begin{equation}
ds^{2}=-(dx^{0})^{2}+\delta_{ij}dx^{i}dx^{j}+O(|x^{j}|^{2})dx^{\alpha}dx^{\beta}.\label{eq: metrica local lorentz}\end{equation}

The effect of the gravitational wave on test masses is described by
the equation

\begin{equation}
\ddot{x^{i}}=-\widetilde{R}_{0k0}^{i}x^{k},\label{eq: deviazione geodetiche}\end{equation}
which is the equation for geodesic deviation in this frame.

Thus, to study the effect of the massive gravitational wave on test
masses, $\widetilde{R}_{0k0}^{i}$ has to be computed in the proper
reference frame of the local observer. But, because the linearized
Riemann tensor $\widetilde{R}_{\mu\nu\rho\sigma}$ is invariant under
gauge transformations \cite{key-12,key-14,key-16,key-19}, it can
be directly computed from eq. (\ref{eq: perturbazione scalare}). 

From \cite{key-19} it is:

\begin{equation}
\widetilde{R}_{\mu\nu\rho\sigma}=\frac{1}{2}\{\partial_{\mu}\partial_{\beta}h_{\alpha\nu}+\partial_{\nu}\partial_{\alpha}h_{\mu\beta}-\partial_{\alpha}\partial_{\beta}h_{\mu\nu}-\partial_{\mu}\partial_{\nu}h_{\alpha\beta}\},\label{eq: riemann lineare}\end{equation}

that, in the case eq. (\ref{eq: perturbazione scalare}), begins

\begin{equation}
\widetilde{R}_{0\gamma0}^{\alpha}=\frac{1}{16}\{\partial^{\alpha}\partial_{0}h_{g}\eta_{0\gamma}+\partial_{0}\partial_{\gamma}h_{g}\delta_{0}^{\alpha}-\partial^{\alpha}\partial_{\gamma}h_{g}\eta_{00}-\partial_{0}\partial_{0}h_{g}\delta_{\gamma}^{\alpha}\};\label{eq: riemann lin scalare}\end{equation}

\begin{equation}
-\partial_{0}\partial_{0}h_{g}\delta_{\gamma}^{\alpha}=\begin{array}{ccc}
-\partial_{z}^{2}h_{g} & for & \alpha=\gamma\end{array}.\label{eq: calcoli4}\end{equation}

The computation has been performed in \cite{key-15}, obtaining

\begin{equation}
\begin{array}{c}
\widetilde{R}_{010}^{1}=-\frac{1}{16}\ddot{h}_{g}\\
\\\widetilde{R}_{020}^{2}=-\frac{1}{16}\ddot{h}_{g}\\
\\\widetilde{R}_{030}^{3}=\frac{1}{16}m_{g}^{2}h_{g}.\end{array}\label{eq: componenti riemann}\end{equation}

The third of eqs. (\ref{eq: componenti riemann}) shows that the field
is not transversal. 

Infact, using eq. (\ref{eq: deviazione geodetiche}) it results

\begin{equation}
\ddot{x}=\frac{1}{16}\ddot{h}_{g}x,\label{eq: accelerazione mareale lungo x}\end{equation}

\begin{equation}
\ddot{y}=\frac{1}{16}\ddot{h}_{g}y\label{eq: accelerazione mareale lungo y}\end{equation}

and 

\begin{equation}
\ddot{z}=-\frac{1}{16}m_{g}^{2}h_{g}(t-vz)z.\label{eq: accelerazione mareale lungo z}\end{equation}

Then the effect of the mass is the generation of a \textit{longitudinal}
force (in addition to the transverse one). 

For a better understanding of this longitudinal force, in \cite{key-14}
the effect on test masses in the context of the geodesic deviation
has been analysed.

Following \cite{key-14} one puts\begin{equation}
\widetilde{R}_{0j0}^{i}=\frac{1}{16}\left(\begin{array}{ccc}
-\partial_{t}^{2} & 0 & 0\\
0 & -\partial_{t}^{2} & 0\\
0 & 0 & m_{g}^{2}\end{array}\right)h_{g}(t-vz)=-\frac{1}{16}T_{ij}\partial_{t}^{2}h_{g}+\frac{1}{16}L_{ij}m_{g}^{2}h_{g}.\label{eq: eqs}\end{equation}

Here the transverse projector with respect to the direction of propagation
of the GW $\widehat{n}$, defined by

\begin{equation}
T_{ij}=\delta_{ij}-\widehat{n}_{i}\widehat{n}_{j},\label{eq: Tij}\end{equation}

and the longitudinal projector defined by

\begin{equation}
L_{ij}=\widehat{n}_{i}\widehat{n}_{j}\label{eq: Lij}\end{equation}

have been used \cite{key-14}. In this way the geodesic deviation
equation (\ref{eq: deviazione geodetiche}) can be rewritten like

\begin{equation}
\frac{d^{2}}{dt^{2}}x_{i}=\frac{1}{16}\partial_{t}^{2}h_{g}T_{ij}x_{j}-\frac{1}{16}m_{g}^{2}h_{g}L_{ij}x_{j}.\label{eq: TL}\end{equation}

Thus it appears clear what was claimed in previous discussion: the
effect of the mass present in the GW generates a longitudinal force
proportional to $m_{g}^{2}$ which is in addition to the transverse
one. But if $v(\omega)\rightarrow1$ in eq. (\ref{eq: velocita' di gruppo 2})
we get $m_{g}\rightarrow0$, and the longitudinal force vanishes.
Thus it is clear that the longitudinal mode arises from the fact that
the GW does no propagate at the speed of light.

In \cite{key-14} it has also been analized the dectability of the
polarization (\ref{eq: hg}) computing the pattern function of a detector
to this massive component. One has to recall that it is possible to
associate to a detector a \textit{detector tensor} that, for an interferometer
with arms along the $\hat{u}$ e $\hat{v}$ directions with respect
the propagating gravitational wave (see figure 1), is defined by \cite{key-2,key-14,key-17}\begin{equation}
D^{ij}\equiv\frac{1}{2}(\hat{v}^{i}\hat{v}^{j}-\hat{u}^{i}\hat{u}^{j}).\label{eq: definizione D}\end{equation}

\begin{figure}
\includegraphics{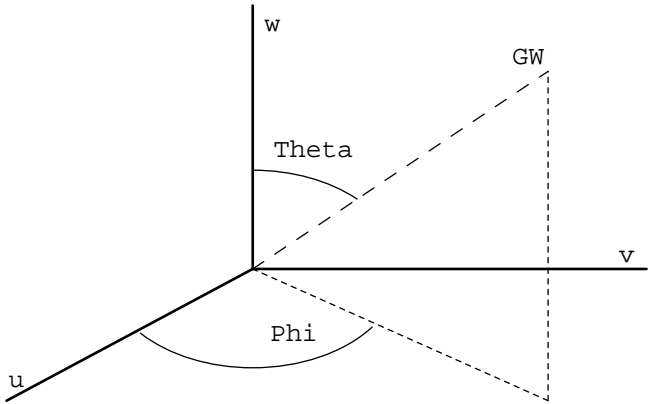}

\caption{a GW propagating from an arbitrary direction}

\end{figure}
 If the detector is an interferometer \cite{key-1,key-2,key-3,key-4,key-5,key-6,key-7,key-8},
the signal induced by a gravitational wave of a generic polarization,
here labelled with $s(t),$ is the phase shift, which is proportional
to \cite{key-2,key-14,key-17}

\begin{equation}
s(t)\sim D^{ij}\widetilde{R}_{i0j0}\label{eq: legame onda-output}\end{equation}

and, using equations (\ref{eq: eqs}), one gets

\begin{equation}
s(t)\sim-\sin^{2}\theta\cos2\phi.\label{eq: legame onda-output 2}\end{equation}

The angular dependence (\ref{eq: legame onda-output 2}), is different
from the two well known standard ones arising from general relativity
which are, respectively 

\begin{center}
\[
(1+\cos^{2}\theta)\cos2\phi\]
 
\par\end{center}

for the $+$ polarization and

\begin{center}
\[
-\cos\theta\sin2\vartheta\]
 
\par\end{center}

for the $\times$ polarization. 

Thus, in principle, the angular dependence (\ref{eq: legame onda-output 2})
could be used to discriminate among the bimetric theory and general
relativity, if present or future detectors will achieve a high sensitivity.

\section{The longitudinal response function}

But there is a problem. The function (\ref{eq: legame onda-output 2})
is not the general form of the response function, but it is only a
good approximation for long wavelengths (i.e. the wavelength of the
wave is much larger than the linear dimension of the interferometer)
\cite{key-2,key-3,key-12,key-17}. Now the full frequency dependent
response function will be computed. For a sake of semplicity we will
consider the case of a massive gravitational way propagating in a
direction parallel to one arm of the interferometer. 

We first performe the computation in the gauge (\ref{eq: metrica puramente scalare}),
which is not the gauge of the local observer in which previous computations
have been performed. Eq. (\ref{eq: metrica puramente scalare}) can
be rewritten as \cite{key-18}

\begin{equation}
(\frac{dt}{d\tau})^{2}-(\frac{dx}{d\tau})^{2}-(\frac{dy}{d\tau})^{2}-(\frac{dz}{d\tau})^{2}=\frac{1}{(1+\frac{1}{8}h_{g})},\label{eq: Sh2}\end{equation}

where $\tau$ is the proper time of the test masses.

From eqs. (\ref{eq: metrica puramente scalare}) and (\ref{eq: Sh2})
the geodesic equations of motion for test masses (i.e. the beam-splitter
and the mirrors of the interferometer), can be obtained\begin{equation}
\begin{array}{ccc}
\frac{d^{2}x}{d\tau^{2}} & = & 0\\
\\\frac{d^{2}y}{d\tau^{2}} & = & 0\\
\\\frac{d^{2}t}{d\tau^{2}} & = & \frac{1}{2}\frac{\partial_{t}(1+\frac{1}{8}h_{g})}{(1+\frac{1}{8}h_{g})^{2}}\\
\\\frac{d^{2}z}{d\tau^{2}} & = & -\frac{1}{2}\frac{\partial_{z}(1+\frac{1}{8}h_{g})}{(1+\frac{1}{8}h_{g})^{2}}.\end{array}\label{eq: geodetiche Corda}\end{equation}

Note: equations (\ref{eq: geodetiche Corda}) are different from equations
(\ref{eq: accelerazione mareale lungo x}), (\ref{eq: accelerazione mareale lungo y})
and (\ref{eq: accelerazione mareale lungo z}) because the gauge (\ref{eq: metrica puramente scalare})
is not the gauge of the local observer where equations (\ref{eq: accelerazione mareale lungo x}),
(\ref{eq: accelerazione mareale lungo y}) and (\ref{eq: accelerazione mareale lungo z})
have been performed. The gauge-invariance between the two gauges will
be shown in Section 4. The first and the second of eqs. (\ref{eq: geodetiche Corda})
can be immediately integrated obtaining

\begin{equation}
\frac{dx}{d\tau}=C_{1}=const.\label{eq: integrazione x}\end{equation}

\begin{equation}
\frac{dy}{d\tau}=C_{2}=const.\label{eq: integrazione x}\end{equation}

In this way eq. (\ref{eq: Sh2}) becomes\begin{equation}
(\frac{dt}{d\tau})^{2}-(\frac{dz}{d\tau})^{2}=\frac{1}{(1+\frac{1}{8}h_{g})}.\label{eq: Ch3}\end{equation}

If we assume that test masses are at rest initially we get $C_{1}=C_{2}=0$.
Thus we see that, even if the GW arrives at test masses, we do not
have motion of test masses within the $x-y$ plane in this gauge.
We could understand this directly from eq. (\ref{eq: metrica puramente scalare})
because the absence of the $x$ and of the $y$ dependences in the
metric implies that test masses momentum in these directions (i.e.
$C_{1}$ and $C_{2}$ respectively) is conserved. This results, for
example, from the fact that in this case the $x$ and $y$ coordinates
do not esplicitly enter in the Hamilton-Jacobi equation for a test
mass in a gravitational field \cite{key-16,key-21}. 

Now we will see that, in presence of the GW, we have motion of test
masses in the $z$ direction which is the direction of the propagating
wave. An analysis of eqs. (\ref{eq: geodetiche Corda}) shows that,
to simplify equations, we can introduce the retarded and advanced
time coordinates ($a,b$):

\begin{equation}
\begin{array}{c}
a=t-vz\\
\\b=t+vz.\end{array}\label{eq: ret-adv}\end{equation}

From the third and the fourth of eqs. (\ref{eq: geodetiche Corda})
we have

\begin{equation}
\frac{d}{d\tau}\frac{da}{d\tau}=\frac{\partial_{b}[1+\frac{1}{8}h_{g}(a)]}{(1+\frac{1}{8}h_{g}(a))^{2}}=0.\label{eq: t-z t+z}\end{equation}

This equation can be integrated obtaining

\begin{equation}
\frac{da}{d\tau}=\alpha,\label{eq: t-z}\end{equation}

where $\alpha$ is an integration constant. From eqs. (\ref{eq: Ch3})
and (\ref{eq: t-z}), we also get

\begin{equation}
\frac{db}{d\tau}=\frac{\beta}{1+\frac{1}{8}h_{g}}\label{eq: t+z}\end{equation}

where $\beta\equiv\frac{1}{\alpha}$, and

\begin{equation}
\tau=\beta a+\gamma,\label{eq: tau}\end{equation}

where the integration constant $\gamma$ correspondes simply to the
retarded time coordinate translation $a$. Thus, without loss of generality,
we can put it equal to zero. Now let us see what is the meaning of
the other integration constant $\beta.$ We can write the equation
for $z$ from eqs. (\ref{eq: t-z}) and (\ref{eq: t+z}):

\begin{equation}
\frac{dz}{d\tau}=\frac{1}{2\beta}(\frac{\beta^{2}}{1+\frac{1}{8}h_{g}}-1).\label{eq: z}\end{equation}

When it is $h_{g}=0$ (i.e. before the GW arrives at the test masses)
eq. (\ref{eq: z}) becomes\begin{equation}
\frac{dz}{d\tau}=\frac{1}{2\beta}(\beta^{2}-1).\label{eq: z ad h nullo}\end{equation}

But this is exactly the initial velocity of the test mass, then we
have to choose $\beta=1$ because we suppose that test masses are
at rest initially. This also imply $\alpha=1$.

To find the motion of a test mass in the $z$ direction we see that
from eq. (\ref{eq: tau}) we have $d\tau=da$, while from eq. (\ref{eq: t+z})
we have $db=\frac{d\tau}{1+\frac{1}{8}h_{g}}$. Because it is $vz=\frac{b-a}{2}$
we obtain

\begin{equation}
dz=\frac{1}{2v}(\frac{d\tau}{1+\frac{1}{8}h_{g}}-da),\label{eq: dz}\end{equation}

which can be integrated as

\begin{equation}
\begin{array}{c}
z=z_{0}+\frac{1}{2v}\int(\frac{da}{1+\frac{1}{8}h_{g}}-da)=\\
\\=z_{0}-\frac{1}{2v}\int_{-\infty}^{t-vz}\frac{\frac{1}{8}h_{g}(a)}{1+\frac{1}{8}h_{g}(a)}da,\end{array}\label{eq: moto lungo z}\end{equation}

where $z_{0}$ is the initial position of the test mass. Now the displacement
of the test mass in the $z$ direction can be written as

\begin{equation}
\begin{array}{c}
\Delta z=z-z_{0}=-\frac{1}{2v}\int_{-\infty}^{t-vz_{0}-v\Delta z}\frac{\frac{1}{8}h_{g}(a)}{1+\frac{1}{8}h_{g}(a)}da\\
\\\simeq-\frac{1}{2v}\int_{-\infty}^{t-vz_{0}}\frac{\frac{1}{8}h_{g}(a)}{1+\frac{1}{8}h_{g}(a)}da.\end{array}\label{eq: spostamento lungo z}\end{equation}
We can also rewrite our results in function of the time coordinate
$t$:

\begin{equation}
\begin{array}{ccc}
x(t) & = & x_{0}\\
\\y(t) & = & y_{0}\\
\\z(t) & = & z_{0}-\frac{1}{2v}\int_{-\infty}^{t-vz_{0}}\frac{\frac{1}{8}h_{g}(a)}{1+\frac{1}{8}h_{g}(a)}d(a)\\
\\\tau(t) & = & t-vz(t),\end{array}\label{eq: moto gauge Corda}\end{equation}

Calling $l$ and $L+l$ the unperturbed positions of the beam-splitter
and of the mirror and using the third of eqs. (\ref{eq: moto gauge Corda})
the varying position of the beam-splitter and of the mirror are given
by

\begin{equation}
\begin{array}{c}
z_{BS}(t)=l-\frac{1}{2v}\int_{-\infty}^{t-vl}\frac{\frac{1}{8}h_{g}(a)}{1+\frac{1}{8}h_{g}(a)}d(a)\\
\\z_{M}(t)=L+l-\frac{1}{2v}\int_{-\infty}^{t-v(L+l)}\frac{\frac{1}{8}h_{g}(a)}{1+\frac{1}{8}h_{g}(a)}d(a)\end{array}\label{eq: posizioni}\end{equation}

But we are interested in variations in the proper distance (time)
of test masses, thus, in correspondence of eqs. (\ref{eq: posizioni}),
using the fourth of eqs. (\ref{eq: moto gauge Corda}) we get\begin{equation}
\begin{array}{c}
\tau_{BS}(t)=t-vl-\frac{1}{2}\int_{-\infty}^{t-vl}\frac{\frac{1}{8}h_{g}(a)}{1+\frac{1}{8}h_{g}(a)}d(a)\\
\\\tau_{M}(t)=t-vL-vl-\frac{1}{2}\int_{-\infty}^{t-v(L+l)}\frac{\frac{1}{8}h_{g}(a)}{1+\frac{1}{8}h_{g}(a)}d(a).\end{array}\label{eq: posizioni 2}\end{equation}

Then the total variation of the proper time is given by

\begin{equation}
\bigtriangleup\tau(t)=\tau_{M}(t)-\tau_{BS}(t)=vL-\frac{1}{2}\int_{t-vl}^{t-v(L+l)}\frac{\frac{1}{8}h_{g}(a)}{1+\frac{1}{8}h_{g}(a)}d(a).\label{eq: time}\end{equation}

In this way, recalling that in the used units the unperturbed proper
distance (time) is $T=L$, the difference between the total variation
of the proper time in presence and the total variation of the proper
time in absence of the GW is \begin{equation}
\delta\tau(t)\equiv\bigtriangleup\tau(t)-L=-L(v+1)-\frac{1}{2}\int_{t-vl}^{t-v(L+l)}\frac{\frac{1}{8}h_{g}(a)}{1+\frac{1}{8}h_{g}(a)}d(a).\label{eq: time variation}\end{equation}

This quantity can be computed in the frequency domain, defining the
Fourier transform of $h_{g}$ as \begin{equation}
\widetilde{h}_{g}(\omega)=\int_{-\infty}^{\infty}dt\textrm{ }h_{g}(t)\exp(i\omega t).\label{eq: trasformata di fourier}\end{equation}

and using the translation and derivation Fourier theorems, obtaining\begin{equation}
\begin{array}{c}
\delta\widetilde{\tau}(\omega)=\{L(1-v^{2})\exp[i\omega L(1+v)]+\frac{L}{2\omega L(v^{2}-1)^{2}}\\
\\{}[\exp[2i\omega L](v+1)^{3}(-2i+\omega L(v-1)+2L\exp[i\omega L(1+v)]\\
\\(6iv+2iv^{3}-\omega L+\omega Lv^{4})+L(v+1)^{3}(-2i+\omega L(v+1))\}\frac{1}{8}\widetilde{h}_{g}.\end{array}\label{eq: segnale totale lungo z}\end{equation}

A {}``signal'' can be also defined:

\begin{equation}
\begin{array}{c}
\widetilde{S}(\omega)\equiv\frac{\delta\widetilde{\tau}(\omega)}{L}=\{(1-v^{2})\exp[i\omega L(1+v_{G})]+\frac{1}{2\omega L(v^{2}-1)^{2}}\\
\\{}[\exp[2i\omega L](v+1)^{3}(-2i+\omega L(v-1)+2\exp[i\omega L(1+v)]\\
\\(6iv+2iv^{3}-\omega L+\omega Lv^{4})+L(v+1)^{3}(-2i+\omega L(v+1))\}\frac{1}{8}\widetilde{h}_{g}.\end{array}\label{eq: sig}\end{equation}

Then the function \begin{equation}
\begin{array}{c}
\Upsilon_{l}(\omega)\equiv(1-v^{2})\exp[i\omega L(1+v_{G})]+\frac{1}{2\omega L(v^{2}-1)^{2}}\\
\\{}[\exp[2i\omega L](v+1)^{3}(-2i+\omega L(v-1)+2\exp[i\omega L(1+v)]\\
\\(6iv+2iv^{3}-\omega L+\omega Lv^{4})+(v+1)^{3}(-2i+\omega L(v+1))],\end{array}\label{eq: risposta totale lungo z due}\end{equation}

is the response function of an arm of our interferometer located in
the $z$-axis, due to the longitudinal component of the massive gravitational
wave arising from the bimetric theory of gravity and propagating in
the same direction of the axis.

For $v\rightarrow1$ it is $\Upsilon_{l}(\omega)\rightarrow0$. 

In figures 2, 3 and 4 are shown the response functions (\ref{eq: risposta totale lungo z due})
for an arm of the Virgo interferometer ($L=3Km$) for $v=0.1$ (non-relativistic
case), $v=0.9$ (relativistic case) and $v=0.999$ (ultra-relativistic
case). We see that in the non-relativistic case the signal is stronger
as it could be expected (for $m_{g}\rightarrow0$ we expect$\Upsilon_{l}(\omega)\rightarrow0$).
In figures 5, 6, and 7 the same response functions are shown for the
Ligo interferometer ($L=4Km$).

\begin{figure}
\includegraphics{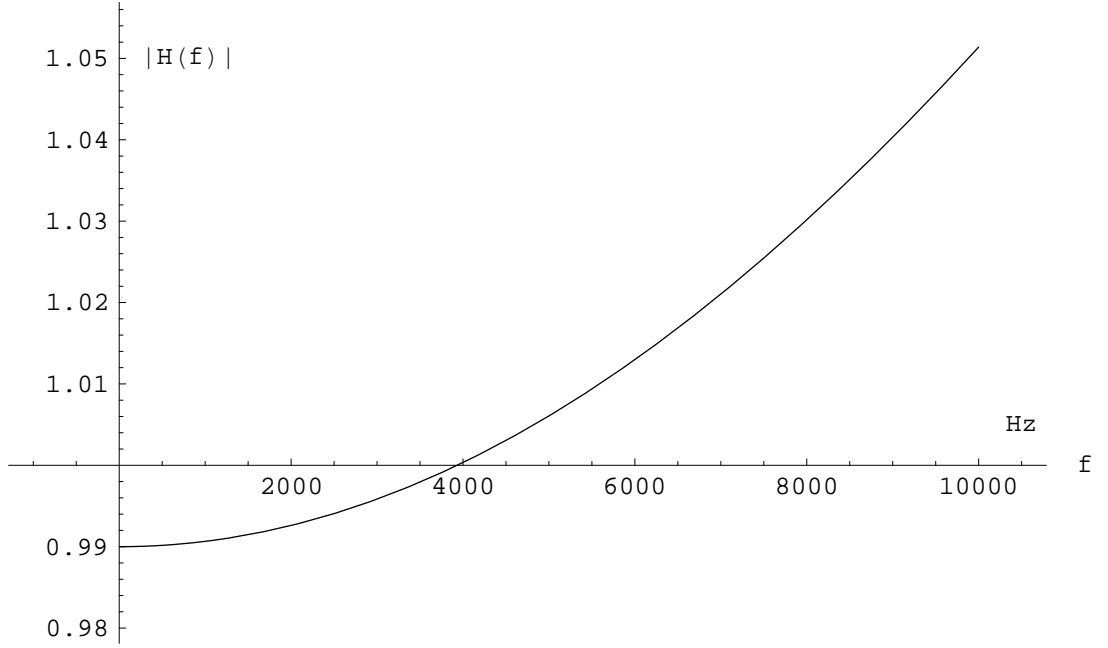}

\caption{the absolute value of the longitudinal response function (\ref{eq: risposta totale lungo z due})
of the Virgo interferometer ($L=3Km$) to a GW arising from the bimetric
theory of gravity and propagating with a speed of $0.1c$ (non relativistic
case). }

\end{figure}
\begin{figure}
\includegraphics{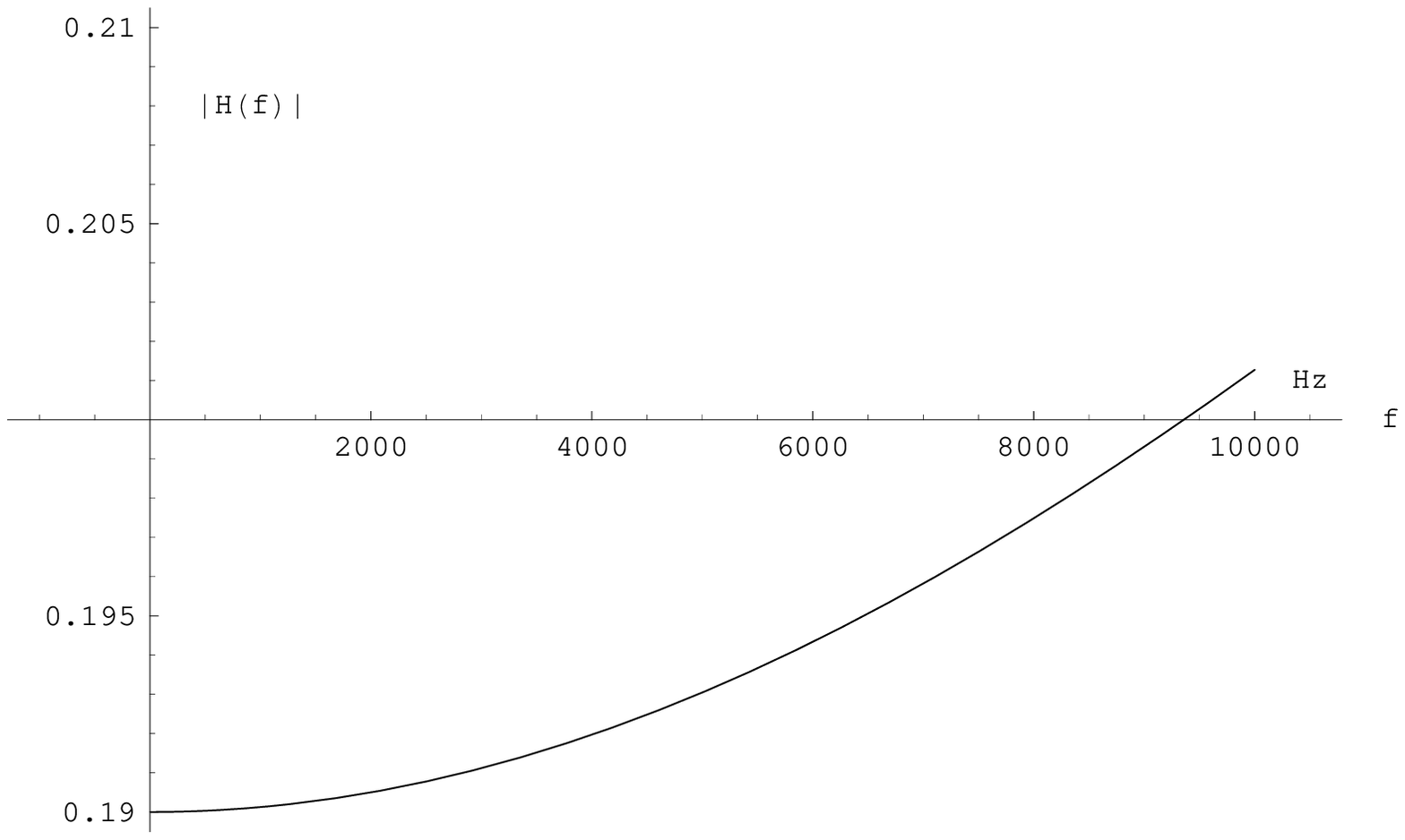}

\caption{the absolute value of the longitudinal response function (\ref{eq: risposta totale lungo z due})
of the Virgo interferometer ($L=3Km$) to a GW arising from the bimetric
theory of gravity and propagating with a speed of $0.9$ (relativistic
case). }

\end{figure}
\begin{figure}
\includegraphics{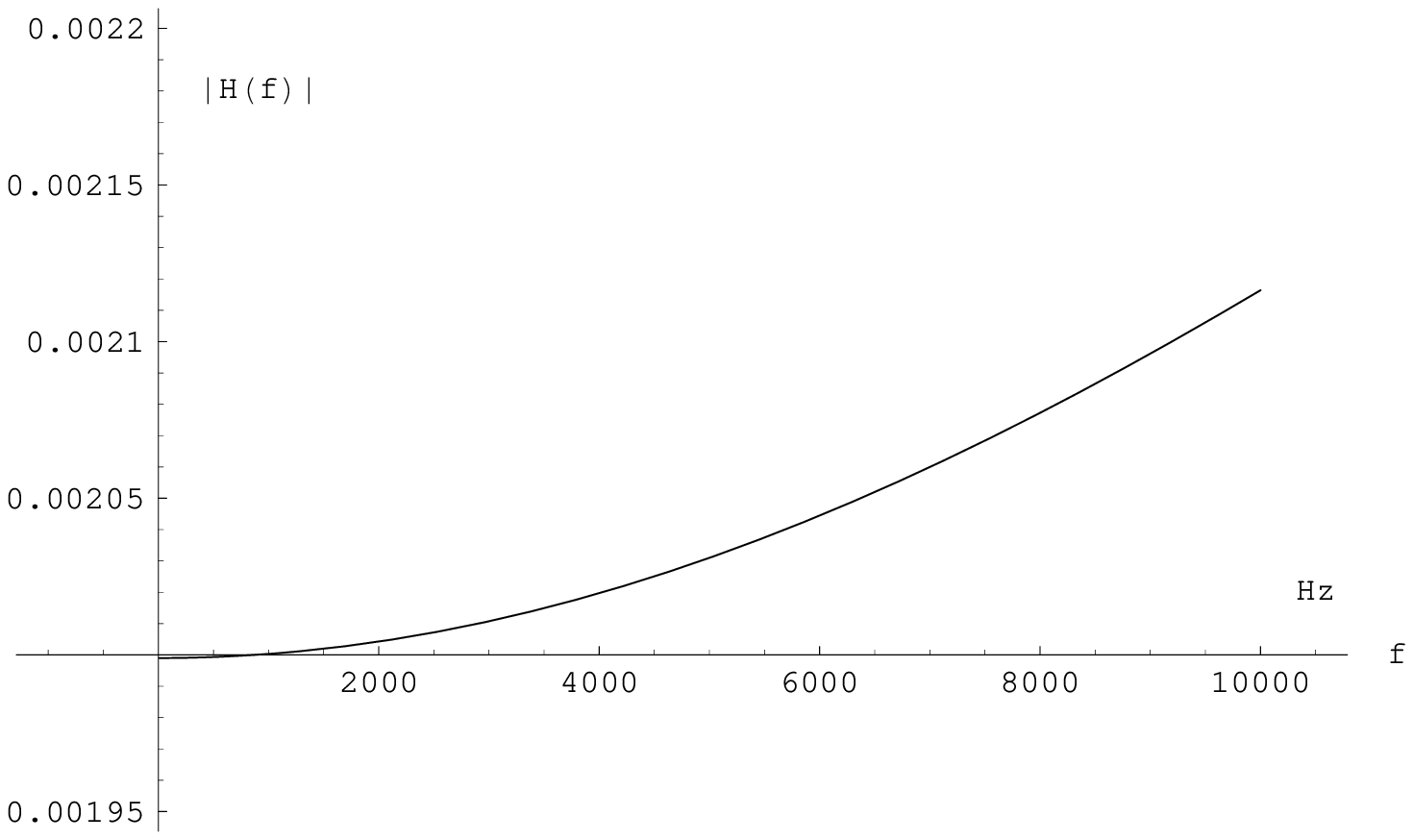}

\caption{the absolute value of the longitudinal response function (\ref{eq: risposta totale lungo z due})
of the Virgo interferometer ($L=3Km$) to a GW arising from the bimetric
theory of gravity and propagating with a speed of $0.999$ (ultra
relativistic case). }

\end{figure}
\begin{figure}
\includegraphics{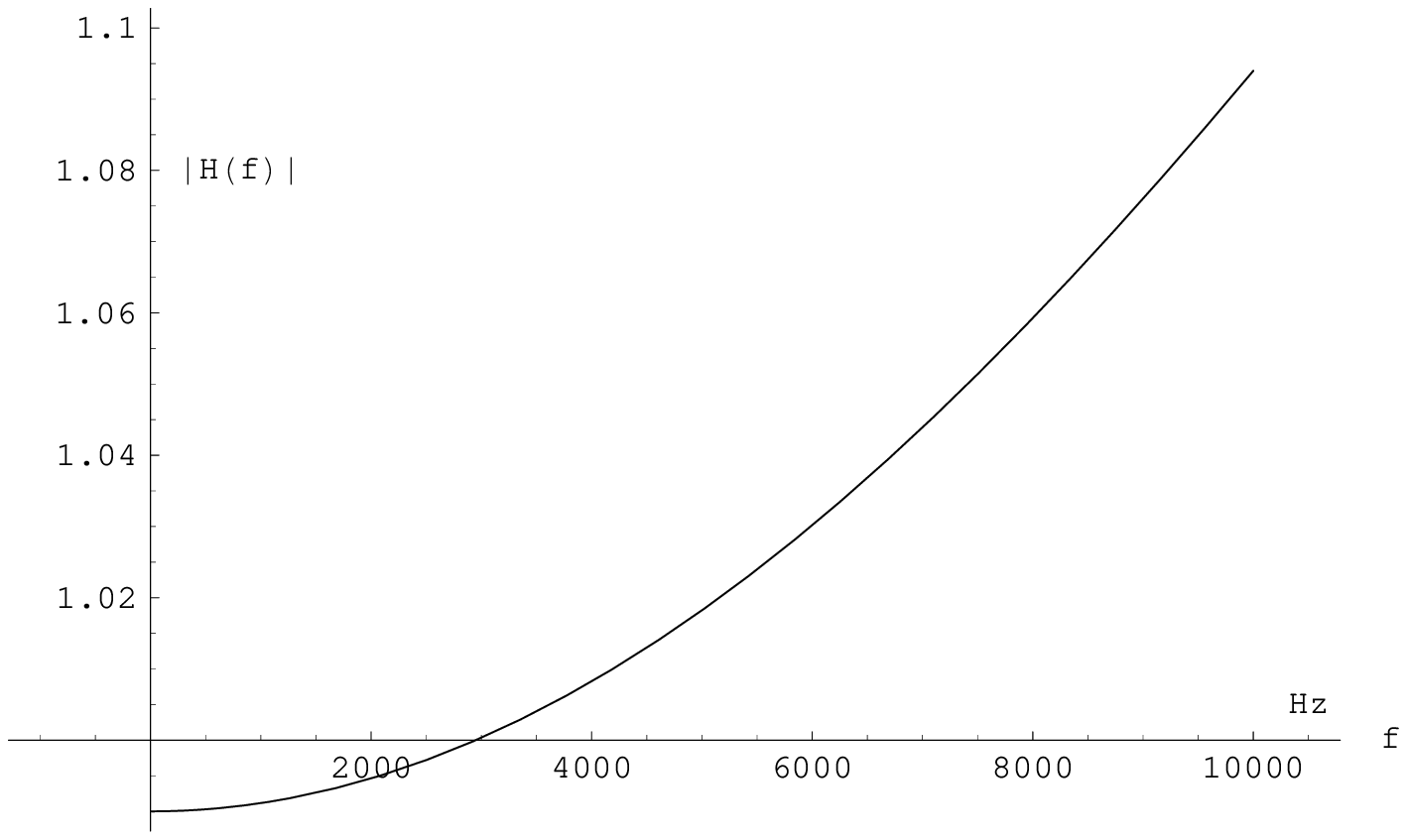}

\caption{the absolute value of the longitudinal response function (\ref{eq: risposta totale lungo z due})
of the LIGO interferometer ($L=4Km$) to a GW arising from the bimetric
theory of gravity and propagating with a speed of $0.1c$ (non relativistic
case). }

\end{figure}
\begin{figure}
\includegraphics{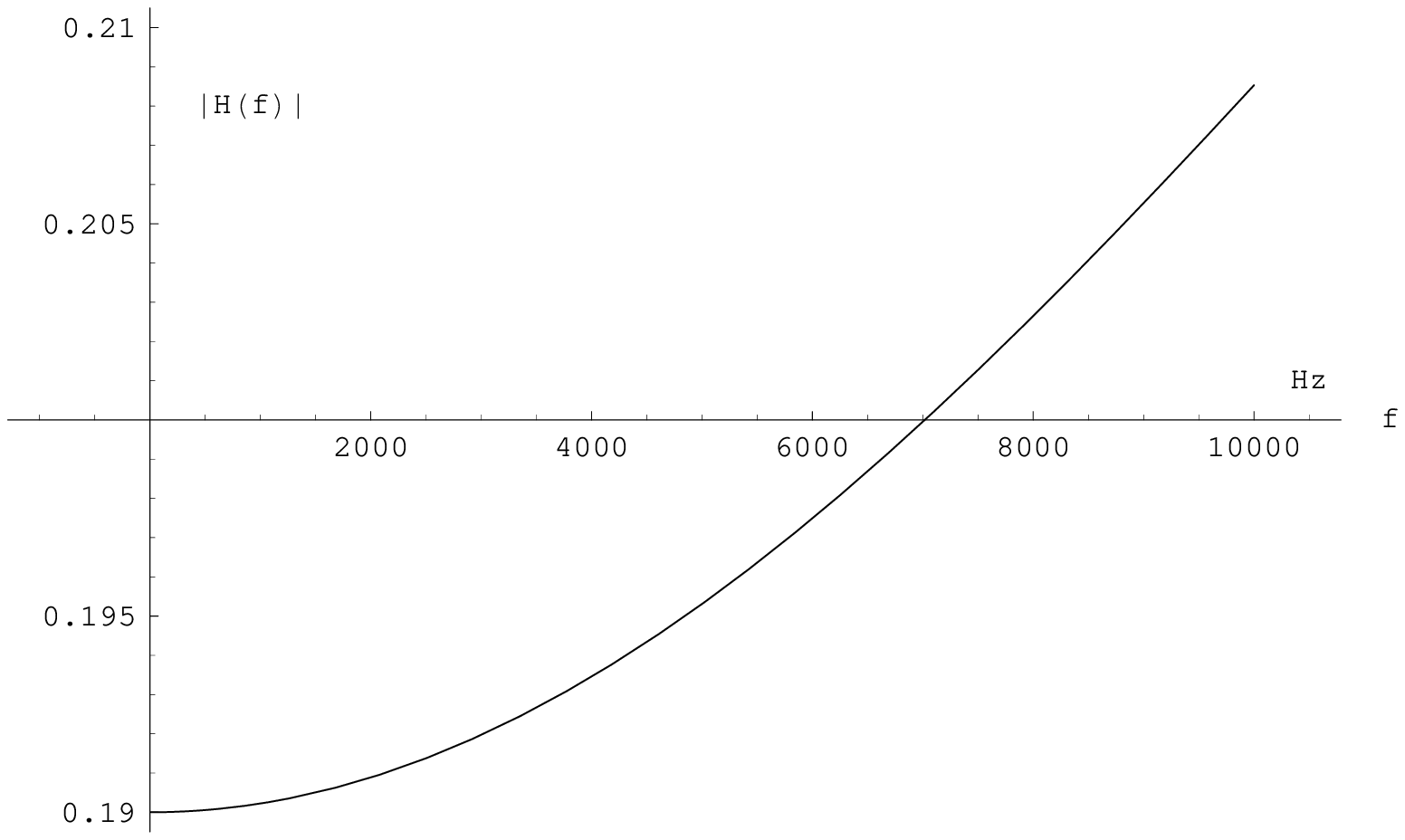}

\caption{the absolute value of the longitudinal response function (\ref{eq: risposta totale lungo z due})
of the LIGO interferometer ($L=4Km$) to a GW arising from the bimetric
theory of gravity and propagating with a speed of $0.9c$ (relativistic
case). }

\end{figure}
\begin{figure}
\includegraphics{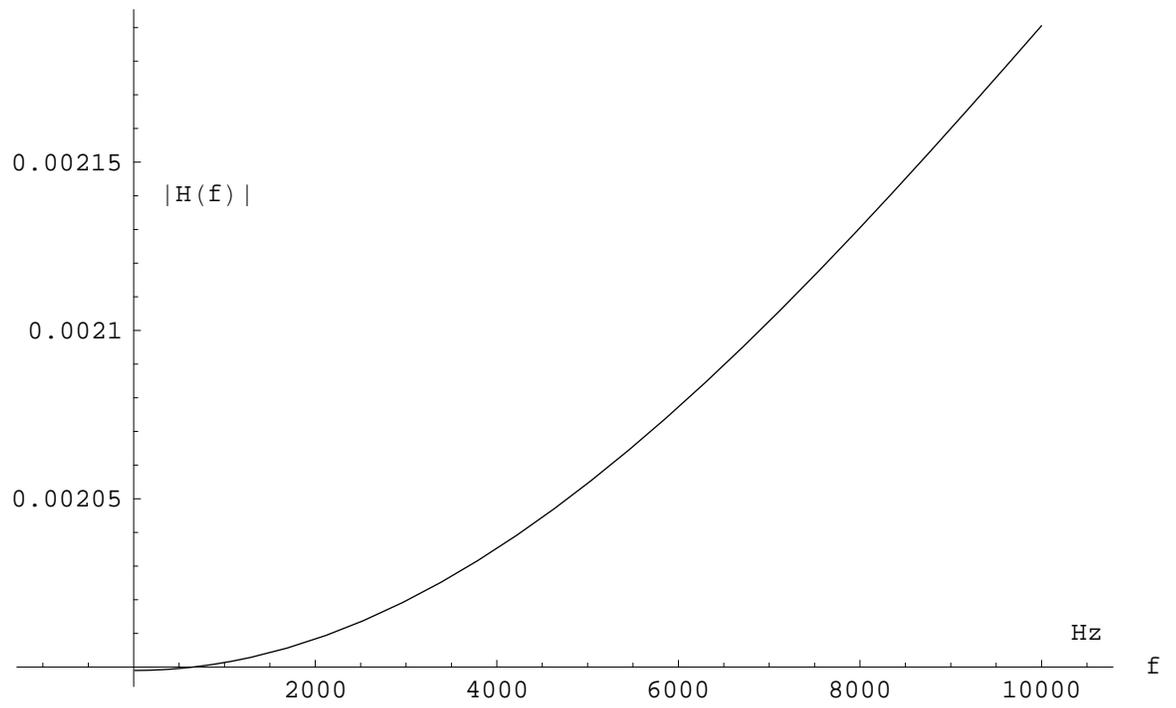}

\caption{the absolute value of the longitudinal response function of the LIGO
interferometer ($L=4Km$) to a GW arising from the bimetric theory
of gravity and propagating with a speed of $0.999c$ (ultra relativistic
case). }

\end{figure}

\section{Gauge invariance of the longitudinal response function}

For a sake of completeness, now the gauge invariance of the longitudinal
response function between the gauge (\ref{eq: metrica puramente scalare})
and the gauge of the local obserer will be shown.

Equations (\ref{eq: accelerazione mareale lungo x}), (\ref{eq: accelerazione mareale lungo y})
and (\ref{eq: accelerazione mareale lungo z}) give the tidal acceleration
of the test mass caused by the gravitational wave respectly in the
$x$ direction, in the $y$ direction and in the $z$ direction \cite{key-16,key-20}.

Equivalently we can say that there is a gravitational potential \cite{key-16,key-19,key-20}:

\begin{equation}
V(\overrightarrow{r},t)=-\frac{1}{32}\ddot{h_{g}}(t-\frac{z}{v})[x^{2}+y^{2}]+\frac{1}{16}m_{g}^{2}2\int_{0}^{z}h_{g}(t-vz)wdw,\label{eq:potenziale in gauge Lorentziana generalizzato}\end{equation}

which generates the tidal forces, and that the motion of the test
mass is governed by the Newtonian equation

\begin{equation}
\ddot{\overrightarrow{r}}=-\bigtriangledown V.\label{eq: Newtoniana}\end{equation}

To obtain the longitudinal component of the gravitational wave the
solution of eq. (\ref{eq: accelerazione mareale lungo z}) has to
be found. 

For this goal the perturbation method can be used \cite{key-16,key-20}.
A function of time for a fixed $z$, $\psi(t-vz)$,  can be defined
\cite{key-16}, for which it is

\begin{equation}
\ddot{\psi}(t-vz)\equiv h_{g}(t-vz)\label{eq: definizione di psi}\end{equation}

(note: the most general definition is $\psi(t-vz)+a(t-vz)+b$, but,
assuming only small variatons in the positions of the test masses,
it results $a=b=0$).

In this way it results

\begin{equation}
\delta z(t-vz)=-\frac{1}{16}m_{g}^{2}z_{0}\psi((t-vz).\label{eq: spostamento lungo z}\end{equation}

A feature of the frame of a local observer is the coordinate dependence
of the tidal forces due by gravitational waves which can be changed
with a mere shift of the origin of the coordinate system \cite{key-16,key-20}:

\begin{equation}
x\rightarrow x+x',\textrm{ }y\rightarrow y+y'\textrm{ and }z\rightarrow z+z'.\label{eq: shift coordinate}\end{equation}

The same applies to the test mass displacements, in the $z$ direction,
eq. (\ref{eq: spostamento lungo z}). This is an indication that the
coordinates of a local observer are not simple as they could seem
\cite{key-16,key-20}. 

Now, let us consider the relative motion of test masses. A good way
to analyze variations in the proper distance (time) of test masses
is by means of {}``bouncing photons'' (see refs. \cite{key-16,key-20}
and figure 8). %
\begin{figure}
\includegraphics{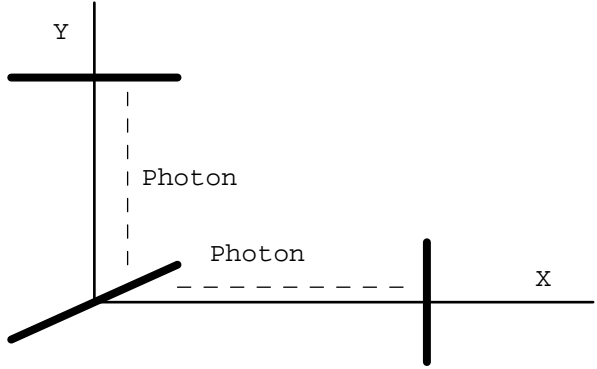}

\caption{photons can be launched from the beam-splitter to be bounced back
by the mirror}

\end{figure}
A photon can be launched from the beam-splitter to be bounced back
by the mirror. It will be assumed that both the beam-splitter and
the mirror are located along the $z$ axis of our coordinate system
(i.e. an arm of the interferometer is in the $z$ direction, which
is the direction of the propagating massive gravitational wave and
of the longitudinal force, see also Figure 9).

\begin{figure}
\includegraphics{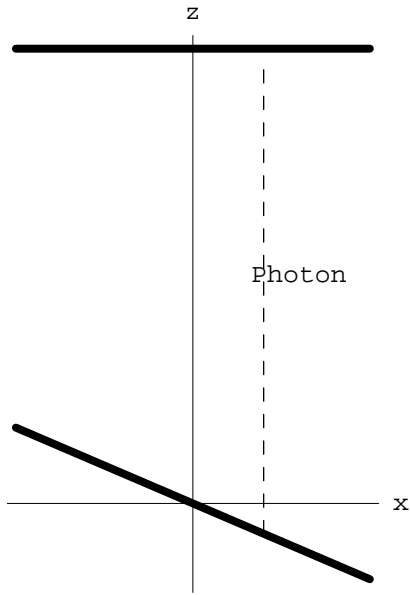}

\caption{the beam splitter and the mirror are located in the direction of the
propagating GW}

\end{figure}

It will be shown that, in the frame of a local observer, two different
effects have to be considered in the calculation of the variation
of the round-trip time for photons, like in \cite{key-16}. Note that
in \cite{key-20} the considered effects were three, but, if we put
the beam splitter in the origin of our coordinate system, the third
effect vanishes \cite{key-16}. 

The unperturbed coordinates for the beam-splitter and the mirror are
$x_{b}=0$ and $x_{m}=L$. Thus, the unperturbed propagation time
between the two masses is

\begin{equation}
T=L.\label{eq: tempo imperturbato}\end{equation}

From eq. (\ref{eq: spostamento lungo z}) it results that the displacements
of the two masses under the influence of the gravitational wave are

\begin{equation}
\delta z_{b}(t)=0\label{eq: spostamento beam-splitter}\end{equation}

and

\begin{equation}
\delta z_{m}(t-vL)=-\frac{1}{16}m_{g}^{2}L\psi(t-vL).\label{eq: spostamento mirror}\end{equation}

In this way, the relative displacement, is

\begin{equation}
\delta L(t)=\delta z_{m}(t-vL)-\delta z_{b}(t)=-\frac{1}{16}m_{g}^{2}L\psi(t-vL),\label{eq: spostamento relativo}\end{equation}

Thus it results

\begin{equation}
\frac{\delta L(t)}{L}=\frac{\delta T(t)}{T}=-\frac{1}{16}m_{g}^{2}\psi(t-vL).\label{eq: strain scalare}\end{equation}

But there is the problem that, for a large separation between the
test masses (in the case of Virgo or LIGO the distance between the
beam-splitter and the mirror is three or four kilometers), the definition
(\ref{eq: spostamento relativo}) for relative displacement becomes
unphysical because the two test masses are taken at the same time
and therefore cannot be in a casual connection \cite{key-16,key-20}.
The correct definitions for our bouncing photon can be written like

\begin{equation}
\delta L_{1}(t)=\delta z_{m}(t-vL)-\delta z_{b}(t-T_{1})\label{eq: corretto spostamento B.S. e M.}\end{equation}

and

\begin{equation}
\delta L_{2}(t)=\delta z_{m}(t-vL-T_{2})-\delta z_{b}(t),\label{eq: corretto spostamento B.S. e M. 2}\end{equation}
where $T_{1}$ and $T_{2}$ are the photon propagation times for the
forward and return trip correspondingly. According to the new definitions,
the displacement of one test mass is compared with the displacement
of the other at a later time to allow for finite delay from the light
propagation. Note that the propagation times $T_{1}$ and $T_{2}$
in eqs. (\ref{eq: corretto spostamento B.S. e M.}) and (\ref{eq: corretto spostamento B.S. e M. 2})
can be replaced with the nominal value $T$ because the test mass
displacements are alredy first order in $h_{g}$. Thus, for the total
change in the distance between the beam splitter and the mirror in
one round-trip of the photon, it is

\begin{equation}
\delta L_{r.t.}(t)=\delta L_{1}(t-T)+\delta L_{2}(t)=2\delta z_{m}(t-vL-T)-\delta z_{b}(t)-\delta z_{b}(t-2T),\label{eq: variazione distanza propria}\end{equation}

and in terms of $\psi$ and of the mass of the gravitational wave:

\begin{equation}
\delta L_{r.t.}(t)=-\frac{1}{8}m_{g}^{2}L\psi(t-vL-T).\label{eq: variazione distanza propria 2}\end{equation}

The change in distance (\ref{eq: variazione distanza propria 2})
leads to changes in the round-trip time for photons propagating between
the beam-splitter and the mirror:

\begin{equation}
\frac{\delta_{1}T(t)}{T}=-\frac{1}{8}m_{g}^{2}\psi(t-vL-T).\label{eq: variazione tempo proprio 1}\end{equation}

In the last calculation (variations in the photon round-trip time
which come from the motion of the test masses inducted by the massive
gravitational wave), it was implicitly assumed that the propagation
of the photon between the beam-splitter and the mirror of our interferometer
is uniform as if it were moving in a flat space-time. But the presence
of the tidal forces indicates that the space-time is curved. As a
result another effect after the previous has to be considered, which
requires spacial separation \cite{key-16,key-20}.

For this effect we consider the interval for photons propagating along
the $z$-axis

\begin{equation}
ds^{2}=g_{00}dt^{2}+dz^{2}.\label{eq: metrica osservatore locale}\end{equation}

The condition for a null trajectory ($ds=0$) gives the coordinate
velocity of the photons

\begin{equation}
v_{f}^{2}\equiv(\frac{dz}{dt})^{2}=1+2V(t,z),\label{eq: velocita' fotone in gauge locale}\end{equation}

which to first order in $h_{g}$ is approximated by

\begin{equation}
v_{f}\approx\pm[1+V(t,z)],\label{eq: velocita fotone in gauge locale 2}\end{equation}

with $+$ and $-$ for the forward and return trip respectively. Knowing
the coordinate velocity of the photon, the propagation time for its
travelling between the beam-splitter and the mirror can be defined:

\begin{equation}
T_{1}(t)=\int_{z_{b}(t-T_{1})}^{z_{m}(t)}\frac{dz}{v_{f}}\label{eq:  tempo di propagazione andata gauge locale}\end{equation}

and

\begin{equation}
T_{2}(t)=\int_{z_{m}(t-T_{2})}^{z_{b}(t)}\frac{(-dz)}{v_{f}}.\label{eq:  tempo di propagazione ritorno gauge locale}\end{equation}

The calculations of these integrals would be complicated because the
boundary $z_{m}(t)$ is changing with time. In fact it is

\begin{equation}
z_{b}(t)=\delta z_{b}(t)=0\label{eq: variazione b.s. in gauge locale}\end{equation}

but

\begin{equation}
z_{m}(t)=L+\delta z_{m}(t).\label{eq: variazione specchio nin gauge locale}\end{equation}

But, to first order in $h_{g}$, this contribution can be approximated
by $\delta L_{2}(t)$ (see eq. (\ref{eq: corretto spostamento B.S. e M. 2})).
Thus, the combined effect of the varying boundary is given by $\delta_{1}T(t)$
in eq. (\ref{eq: variazione tempo proprio 1}). Then only the times
for photon propagation between the fixed boundaries $0$ and $L$
have to be calculated. Such propagation times will be denoted with
$\Delta T_{1,2}$ to distinguish from $T_{1,2}$. In the forward trip,
the propagation time between the fixed limits is

\begin{equation}
\Delta T_{1}(t)=\int_{0}^{L}\frac{dz}{v_{f}(t',z)}\approx T-\int_{0}^{L}V(t',z)dz,\label{eq:  tempo di propagazione andata  in gauge locale}\end{equation}

where $t'$ is the retardation time which corresponds to the unperturbed
photon trajectory: 

\begin{center}
$t'=t-(L-z)$
\par\end{center}

(i.e. $t$ is the time at which the photon arrives in the position
$L$, so $L-z=t-t'$).

Similiary, the propagation time in the return trip is

\begin{equation}
\Delta T_{2}(t)=T-\int_{L}^{0}V(t',z)dz,\label{eq:  tempo di propagazione andata  in gauge locale}\end{equation}

where now the retardation time is given by

\begin{center}
$t'=t-z$.
\par\end{center}

The sum of $\Delta T_{1}(t-T)$ and $\Delta T_{2}(t)$ gives the round-trip
time for photons traveling between the fixed boundaries. Then the
deviation of this round-trip time (distance) from its unperturbed
value $2T$ is

\begin{equation}
\delta_{2}T(t)=\int_{0}^{L}[V(t-2T+z,z)+V(t-z,z)]dz.\label{eq: variazione tempo proprio lungo z 2}\end{equation}

From eqs. (\ref{eq:potenziale in gauge Lorentziana generalizzato})
and (\ref{eq: variazione tempo proprio lungo z 2}) it results:

\begin{equation}
\begin{array}{c}
\delta_{2}T(t)=\frac{1}{2}m_{g}^{2}\int_{0}^{L}[\int_{0}^{z}\frac{1}{8}h_{g}(t-2T+w-vw)wdw+\int_{0}^{z}\frac{1}{8}h_{g}(t-w-vw)wdw]dz=\\
\\=\frac{1}{4}m_{g}^{2}\int_{0}^{L}[\frac{1}{8}h_{g}(t-vz-2T+z)+\frac{1}{8}h_{g}(t-vz-z)]z^{2}dz+\\
\\-\frac{1}{4}m_{g}^{2}\int_{0}^{L}[\int_{0}^{z}\frac{1}{8}h_{g}'(t-2T+w-vw)z^{2}dw+\int_{0}^{z}\frac{1}{8}h_{g}'(t-w-vw)z^{2}dw]dz,\end{array}\label{eq: variazione tempo proprio lungo z 2.2}\end{equation}

Thus the total round-trip proper distance in presence of the massive
gravitational wave is:

\begin{equation}
T=2T+\delta_{1}T+\delta_{2}T.\label{eq: round-trip  totale in gauge locale}\end{equation}

Now, to obtain the interferometer response function of the massive
gravitational wave, the analysis will be transled in the frequency
domine.

Using the Fourier transform of $\psi$ defined from 

\begin{equation}
\tilde{\psi}(\omega)=\int_{-\infty}^{\infty}dt\psi(t)\exp(i\omega t),\label{eq: trasformata di fourier2}\end{equation}
 eq. (\ref{eq: variazione tempo proprio 1}) can be rewritten like:

\begin{equation}
\frac{\delta_{1}\tilde{T}(\omega)}{T}=-\frac{1}{8}m_{g}^{2}\Upsilon_{1}^{*}(\omega)\tilde{\psi}(\omega)\label{eq: fourier 1 lungo z}\end{equation}

with 

\begin{equation}
\Upsilon_{1}^{*}(\omega)=\exp[i\omega(1+v_{G})L].\label{eq: risposta 1 lungo z}\end{equation}
But, from a theorem about Fourier transforms, it is simple to obtain:

\begin{equation}
\tilde{\psi}(\omega)=-\frac{\tilde{h}_{g}(\omega)}{\omega^{2}},\label{eq: Teorema di Fourier}\end{equation}

where the Fourier transform of $h_{g}$ is given by equation (\ref{eq: trasformata di fourier}).

Then it results:

\begin{equation}
\frac{\delta_{1}\tilde{T}(\omega)}{T}=\frac{m_{g}^{2}}{8\omega^{2}}\Upsilon_{1}^{*}(\omega)\tilde{h}_{g}(\omega),\label{eq: delta t su t finale}\end{equation}
and, defining:

\begin{equation}
\Upsilon_{1}\equiv\frac{m_{g}^{2}}{\omega^{2}}\Upsilon_{1}^{*}(\omega)=(1-v^{2})\Upsilon_{1}^{*}(\omega),\label{eq: def.  gamma1}\end{equation}

we obtain:

\begin{equation}
\frac{\delta_{1}\tilde{T}(\omega)}{T}=\frac{1}{8}\Upsilon_{1}(\omega)\tilde{h}_{g}(\omega).\label{eq: delta t su t finale 2}\end{equation}

On the other hand eq. (\ref{eq: variazione tempo proprio lungo z 2.2})
can be rewritten in the frequency space like:

\begin{equation}
+\begin{array}{c}
\delta_{2}\tilde{T}(\omega)=\frac{1}{2\omega(v^{2}-1)^{2}}[\exp[2i\omega L](v+1)^{3}(-2i+\omega L(v-1)+\\
\\2\exp[i\omega L(1+v)](6iv+2iv^{3}-\omega L+\omega Lv^{4})+\\
\\+(v+1)^{3}(-2i+\omega L(v+1))]\frac{\tilde{h}_{g}(\omega)}{8}.\end{array}\label{eq: variazione tempo proprio lungo z 2.3}\end{equation}

Now 

\begin{equation}
\frac{\delta_{2}\tilde{T}(\omega)}{T}=\Upsilon_{2}(\omega)\frac{\tilde{h}_{g}(\omega)}{8},\label{eq: fourier 2 lungo z}\end{equation}

can be put, with 

\begin{equation}
\begin{array}{c}
\Upsilon_{2}(\omega)=\frac{1}{2\omega L(v^{2}-1)^{2}}[\exp[2i\omega L](v+1)^{3}(-2i+\omega L(v-1)+\\
\\2\exp[i\omega L(1+v)](6iv+2iv^{3}-\omega L+\omega Lv^{4})+\\
\\+(v+1)^{3}(-2i+\omega L(v+1))].\end{array}\label{eq: risposta 2 lungo z}\end{equation}
Because it is

\begin{equation}
\Upsilon_{l}(\omega)=\Upsilon_{1}(\omega)+\Upsilon_{2}(\omega),\label{eq: risposta totale lungo z}\end{equation}
from eqs. (\ref{eq: risposta 1 lungo z}), (\ref{eq: def.  gamma1})
and (\ref{eq: risposta 2 lungo z}) it results that the function 

\begin{equation}
\begin{array}{c}
\Upsilon_{l}(\omega)\equiv(1-v^{2})\exp[i\omega L(1+v_{G})]+\frac{1}{2\omega L(v^{2}-1)^{2}}\\
\\{}[\exp[2i\omega L](v+1)^{3}(-2i+\omega L(v-1)+2\exp[i\omega L(1+v)]\\
\\(6iv+2iv^{3}-\omega L+\omega Lv^{4})+(v+1)^{3}(-2i+\omega L(v+1))],\end{array}\label{eq: risposta totale lungo z finale}\end{equation}

is the longitudinal response function of an arm of the interferometer
located in the $z$-axis, due to the longitudinal component of the
massive gravitational wave propagating in the same direction of the
axis, and one can see that equation (\ref{eq: risposta totale lungo z finale})
is equal to equation (\ref{eq: risposta totale lungo z due}).

Thus, we have shown that the longitudinal response function of an
arm of an interferometer located in the $z$-axis is the same in both
the local Lorentz gauge and in the gauge (\ref{eq: metrica puramente scalare}).

\section{Conclusions}

This paper is an integration of previous research on massive gravitational
waves from a bimetric theory of gravity. In the literature about this
issue, it has been shown that massive gravitational waves arising
from such a bimetric theory can generate a longitudinal component
in a particular polarization of the wave \cite{key-14,key-15}. After
a review of previous works, which was due for completeness and for
a better understanding of the analysis, in this paper the longitudinal
response function of interferometers for this particular polarization
of the wave has been computed in two different gauges, showing the
gauge invariance, and in its full frequency dependence, with specific
application to the Virgo and LIGO interferometers.

\section*{Acknowledgements}

I would like to thank  Herman Mosquera Cuesta and Franceso Rubanu
for helpful advices during my work.

\end{document}